\documentclass[aps,twocolumn,nofootinbib]{revtex4-1}
\usepackage[dvipsnames]{xcolor}
\usepackage{slashed}
\usepackage{stmaryrd}
\usepackage{mathrsfs}
\usepackage{yfonts}
\usepackage{rotating}
\usepackage{graphicx}
\usepackage{amsfonts}
\usepackage{amsmath}
\usepackage{mathrsfs}
\usepackage[american]{babel}
\usepackage{wasysym}
\usepackage{slashed}
\usepackage{pgfplots}
\usepackage{braket}
\usepackage{caption}
\usepackage{graphicx}
\usepackage{subfigure}
\usepackage{amssymb}
\usepackage{mathtools}
\usepackage{multirow}
\usepackage{booktabs}

\newcommand{\g}[0]{\gamma}
\newcommand{\al}[0]{\alpha}
\newcommand{\be}[0]{\beta}
\newcommand{\de}[0]{\delta}
\newcommand{\e}[0]{\varepsilon}
\newcommand{\dd}[0]{\partial}
\newcommand{\la}[0]{\lambda}
\newcommand{\om}[0]{\omega}
\newcommand{\si}[0]{\sigma}

\newcommand{\bs}[1]{\textbf{#1}}
\newcommand{\bss}[1]{\boldsymbol{#1}}
\newcommand{\ea}[1]{\begin{align}#1\end{align}}
\newcommand{\eq}[1]{\begin{equation}#1\end{equation}}

\newcommand{\ma}[1]{\mathcal{#1}}

\begin{document}

\title{An exceptional parallel between internal and spacetime symmetries}
\author{Gon\c{c}alo M. Quinta}
\affiliation{Quantum Physics of Information Group, Instituto de Telecomunica\c{c}\~{o}es, Lisbon}
\email{goncalo.quinta@tecnico.ulisboa.pt}

\begin{abstract}
We show that, when expressed in an exceptional Jordan algebra form, the spacetime symmetries and field content of the massless free Dirac theory have a natural parallel to the internal symmetries and field content resulting from a $\textrm{Spin}(10)$ GUT with three light chiral families and vector-like Kaluza-Klein excitations. Within this construction, spacetime has at most ten dimensions.
\end{abstract}

\maketitle

\section{Introduction}

A central difficulty of grand unification is finding a mathematical structure large enough to accommodate the Standard Model (SM), yet constrained enough to explain why nature chose its particular representations. Algebraic approaches offer such constraints.

One special algebra has made scarce yet impactful appearances over the past decade: the exceptional Jordan algebra $J_3(\mathbb{O})$ of $3\times3$ Hermitian octonionic matrices, the only exceptional algebra of quantum observables~\cite{Jordan:1934}. Dubois-Violette proposed it as the algebra of observables of a finite quantum space attached to spacetime, with colour $SU(3)$ arising from the octonion automorphisms that fix an imaginary unit~\cite{DuboisViolette:2016} and, with Todorov, obtained the SM gauge group as the intersection of two maximal subgroups of $F_4$~\cite{Todorov:2018, DuboisViolette:2019}. Boyle related $J_3(\mathbb{C}\otimes\mathbb{O})$ to the SM representations, their minimal left-right symmetric extension and $\mathrm{Spin}(10)$ unification, suggesting three generations relate to triality~\cite{Boyle:2020}. Although these subgroups are now well understood (cf.~\cite{Baez:2026} for a particularly transparent account), why nature singles them out remains unclear. This is one of several algebraic routes to unification spanning decades (cf. \cite{Dixon:1994, Furey:2023, Gresnigt:2026} for some historical context and state-of-the-art).

Containing the SM is not, by itself, what makes the exceptional Jordan algebra special, as many large structures do. What sets it apart is that it already carries ten-dimensional (10D) Minkowski spacetime, since its $2\times2$ blocks are 10D vectors and its off-diagonal entries are the corresponding spinors \cite{Baez:2002}. This octonionic description of 10D vectors and Majorana--Weyl spinors goes back to Kugo and Townsend~\cite{Kugo:1983} and Sudbery~\cite{Sudbery:1984}. Schray first combined both in a $3\times3$ Jordan matrix \cite{Schray:1996}, and the relevant operations are reviewed in \cite{Baez:2010}. This invites a spacetime reading of~\cite{Todorov:2018, DuboisViolette:2019, Boyle:2020}.

In this work we make these connections concrete. The free Dirac theory in four dimensions (4D) can be written as a single determinant in the complexified exceptional Jordan algebra, the complexification being required by the Dirac field itself and its largest extension within this algebra is the 10D Majorana-Weyl Lagrangian. Our central observation concerns the three structures reducing the 10D symmetries to those of 4D spacetime: the Peirce decomposition separating fields from derivatives, the chirality involution and the complex structure of $\mathbb{C}\otimes\mathbb{O}$. None of them is chosen, since all are fixed by the Dirac theory itself, and they yield the Lorentz group times an extra-dimensional family symmetry $SU(4)_F \supset SU(3)_F\times U(1)_F$. Applied to an internal copy of the algebra, the same structures reproduce Boyle's two maximal subgroups \cite{Boyle:2020}, giving the left-right symmetric extension of the SM with $U(1)_{B-L}$ mirroring $U(1)_F$, plus an extra $U(1)$ supplying the internal orbifold phase. On a $T^6/\mathbb{Z}_6$ orbifold, the structure allows either one or three light chiral families and no mirror zero modes, the three-family case including a vector-like fourth family among the Kaluza-Klein (KK) excitations. Manogue and Dray obtained three generations from an octonionic reduction of the 10D massless Dirac equation \cite{Manogue:1999, Dray:1999}, using a preferred imaginary unit and quaternionic subalgebras instead of compactification. Here the quaternionic subalgebra fixes 4D chirality, and the family count follows from the complex structure and the orbifold, paralleling the untwisted sector of $T^6/\mathbb{Z}_N$ string orbifolds, where the three complex planes give rise to three families~\cite{Dixon:1985, Dixon:1986}.

Internal symmetries are thus mirrored by the very structures that make spacetime 4D and the families chiral, as in the Clifford-algebraic approach of~\cite{Quinta:2025}. A 4D scalar (and tensor) sector likewise emerges as the higher-dimensional vector sector accompanying the fermions in a single element of $J_3(\mathbb{C}\otimes\mathbb{O})$.

\section{Dirac QFT in exceptional Jordan form}

We consider the massless free Dirac field $\psi(x)$, with Lagrangian
\eq{\label{DiracL}
\ma{L}_{D} = i\psi^{\dagger}_{\al}(x)(\g^0 \g^{\mu})_{\al\be}\dd_{\mu}\psi_{\be}(x)\,,
}
in a Minkowski spacetime with $\eta^{\mu\nu}=\textrm{diag}(+1,-1,-1,-1)$, with the Einstein notation assumed throughout. Every result below follows from the identity \cite{Quinta:2025}
\eq{\label{g0gmuCl}
i(\g^0 \g^{\mu})_{\al\be} = 2\, \textrm{Tr}\left[\om_{\al} u^{\mu} \om^{\dagger}_{\be} \left(\frac{1+w}{8}\right)\left(\frac{1+ I_e}{2}\right)\right]
}
where $\om_{\al} = (i e_{\al} - e_{9-\al})/2$, $e_{k}$ are the generators of $\ma{C}\ell_{8,0}$ (basis in Appendix~A), 
$u^{\mu} =\exp\left[i \frac{\pi}{2} (\g^0\g^{\mu})_{\al\be} \om^{\al \dagger}\om^{\be}\right] \in \textrm{Spin}(8)$, $I_e = e_{1\ldots8}$ is the pseudoscalar and $w = e_{1234} + e_{1467} - e_{1357} + e_{1256} - e_{3456}  - e_{2457} - e_{2367}$ is a calibration, with $e_{ij\ldots k} \equiv e_i e_j \ldots e_k$. The $\om_{\al}$ and their Hermitian conjugates $\om^{\dagger}_{\al}$ form a Witt basis of $\ma{C}\ell_{8,0}$, $\{\om_{\al},\om^{\dagger}_{\be}\} = \de_{\al\be}$ and $\{\om^{\dagger}_{\al},\om^{\dagger}_{\be}\} = \{\om_{\al},\om_{\be}\} = 0$, resembling the relations of fermionic quantum fields. In fact, the need for a Witt basis is directly tied both to the components of a Dirac spinor and its second quantization, which require exactly four creation and annihilation operators, represented as tensor products of $\om_{\al}$ and $\om^{\dagger}_{\al}$ (cf. \cite{Quinta:2025} for a proof).

In $\mathcal{C}\ell_{8,0}$, an octonion is a vector $a = a^{k} e_k \in \mathbb{R}^8$, with product \cite{Lounesto:2001} ${a \circ b = \braket{a n b (1+w)(1-I_e)}_1}$, where $n=e_8$ is the chosen neutral axis and $\braket{x}_1$ the part of $x$ written as a linear combination of $e_k$. Octonions are spanned by seven elements $\textbf{e}_k$ and a unit $\textbf{1}$, with a multiplication table characterized by its Fano plane (cf. Appendix~A for the Fano plane used here). Here, we identify $\textbf{e}_k=e_k$ $(k=1,\ldots,7)$ and $\textbf{1} = e_8$. In octonionic form, Eq.~(\ref{g0gmuCl}) reads (cf. Appendix~A)
\eq{\label{g0gmuOct}
i(\g^0 \g^{\mu})_{\al\be} = 2 \textrm{Re}[ (\bar{\om}_{\al} \circ \bss{\g}^0_E) \circ (\bss{\g}^{\mu} \circ \om^{\dagger}_{\be})]
}
where $\bar{x} = 2\textrm{Re}[x] - x$ is the octonion conjugate, $\textrm{Re}[x]$ the component along $\bs{1}$, $(\bss{\g}^0,\bss{\g}^1,\bss{\g}^2,\bss{\g}^3) = (-i\bs{e}_5,-\bs{e}_3,\bs{e}_6,\bs{e}_4)$ and $\bss{\g}_E^0 = i \bss{\g}^0 = \bs{e}_5$. The $\om_{\al} \in \mathbb{C} \otimes \mathbb{O}$ are null octonions $(N(\om_{\al})=0)$ and the $\bss{\g}^{\mu}$ generate $\ma{C}\ell_{1,3}$ by left multiplication, $\{\bss{\g}^{\mu},\bss{\g}^{\nu}\}(x) \equiv \bss{\g}^{\mu} \circ (\bss{\g}^{\nu} \circ x) + \bss{\g}^{\nu} \circ (\bss{\g}^{\mu} \circ x) = 2\eta^{\mu\nu} x$. Left-handed spinors, selected by $P_{L}=(1-\g^5)/2$, span $\mathbb{C} \otimes \mathbb{H}$ (cf.  Appendix~B), with the quaternion algebra
\eq{\label{Halgebra}
\mathbb{H} = \textrm{span}\{\bs{1},\bs{e}_1,\bs{e}_2,\bs{e}_7\}\,,
}
and right-handed ones span $\mathbb{C} \otimes \mathbb{H}^{\perp}$, where $\mathbb{H}^{\perp} = \textrm{span}\{\bs{e}_3,\bs{e}_4,\bs{e}_5,\bs{e}_6\}$.

Equation~(\ref{g0gmuOct}) fits naturally into the exceptional Jordan algebra $J_3(\mathbb C\otimes\mathbb O)$ of $3 \times 3$ Hermitian matrices $X(\xi_1,\xi_2,\xi_3,x_1,x_2,x_3)$, 
\eq{
X =
\begin{pmatrix}
\xi_1 & x_3 & \bar{x}_2 \\
\bar{x}_3 & \xi_2 & x_1 \\
x_2 & \bar{x}_1 & \xi_3
\end{pmatrix}\,,
}
with $\xi_k \in \mathbb{C}$, $x_k \in \mathbb{C} \otimes \mathbb{O}$ and product $X \odot Y = (XY+YX)/2$, using entry-wise octonionic multiplication. We use the decomposition
\eq{\label{GenJordan}
X = 
\begin{pmatrix}
V & \Psi \\
\bar{\Psi}^{T} & \xi_3
\end{pmatrix}\,, \quad
V = 
\begin{pmatrix}
\xi_1 & x_3 \\
\bar{x}_3 & \xi_2
\end{pmatrix}\,, \quad
\Psi =
\begin{pmatrix}
\bar{x}_2 \\
x_1
\end{pmatrix}\,,
}
in which the determinant reads \cite{Dray:2010}
\eq{\label{detX}
\det(X) = \xi_3 \det(V) - \textrm{Re}[\bar{\Psi}^{T} \tilde{V} \Psi]
}
where $\tilde{V} = (\textrm{Tr}[V])\,1_{2\times 2} - V$, $\det(V) = \xi_1 \xi_2 - N(x_3)$, $N(x_k) \equiv x_k \circ \bar{x}_k = \bar{x}_k \circ x_k \in \mathbb{C}$ is the norm and $\textrm{Re}[\bar{\Psi}^{T} \tilde{V} \Psi] = \xi_1 N(x_1)+\xi_2 N(x_2) -2\textrm{Re}[x_1 \circ (x_2 \circ x_3)]$ is unambiguous since $\textrm{Re}[(\bar{\Psi}^{T} \tilde{V}) \Psi] = \textrm{Re}[\bar{\Psi}^{T} (\tilde{V} \Psi)]$. As $\textrm{Re}[x_1 \circ (x_2 \circ x_3)]$ is cyclic, Eq.~(\ref{g0gmuOct}) reads $i(\g^0 \g^{\mu})_{\al\be} = 2\textrm{Re}[x_2 \circ (x_3 \circ x_1)]$ with $x_1 = \om^{\dagger}_{\be}$, $x_2 = \bar{\om}_{\al} \circ \bss{\g}_E^0$ and $x_3 = \bss{\g}^{\mu}$. Since $N(x_1)=N(x_2)=0$ and $N(x_3) = N(\bss{\g}^{\mu}) = -\eta^{\mu\mu}$, setting $\xi_3=0$ gives
\eq{\label{g0gmuDet}
i(\g^0\g^{\mu})_{\al\be} = \det(\Xi^{\mu}_{\al\be})
}
with
\eq{\label{Ximualbe}
\Xi^{\mu}_{\al\be} = 
\begin{pmatrix}
\xi_1 & \bss{\g}^{\mu} & -\bss{\g}_E^0 \circ \om_{\al} \\
\bar{\bss{\g}}^{\mu} & \xi_2 & \om^{\dagger}_{\be} \\
\bar{\om}_{\al} \circ \bss{\g}_E^0 & \bar{\om}^{\dagger}_{\be}  & 0
\end{pmatrix}\,.
}
Inserting this in (\ref{DiracL}) gives the exceptionally simple form
\eq{\label{JordanL}
\ma{L}_D = \det(\Xi)
}
with
\eq{\label{Xi4D}
\Xi = X(\xi_1,\xi_2,0,\psi,\psi^{\dagger}\circ \bss{\g}_E^0, \bss{\g}^{\mu}\dd_{\mu})\,,
}
where $\psi=\sum_{\be} \psi_{\be}\om_{\be}^\dagger$, $\psi^\dagger=\sum_{\al}\psi_{\al}^\dagger\bar{\om}_{\al}$ and Eq.~(\ref{JordanL}) is a shorthand for the ordered contraction $\psi^{\dagger}_{\al}\det(\Xi^{\mu}_{\al\be})\dd_{\mu}\psi_{\be}$, since cyclic permutations of $\textrm{Re}[x_2 \circ (x_3 \circ x_1)]$ would misplace the derivative. The variables $\xi_1,\xi_2$ cancel in Eq.~(\ref{JordanL}) but acquire a meaning once $V$ is written as $V(p)=p_A\bss{\si}^A$, with
\eq{
\bss{\si}^{a} =
\begin{pmatrix}
0 & \bss{\g}^{a} \\
\bar{\bss{\g}}^a & 0
\end{pmatrix},\,\,\,\,
\bss{\si}^8 = 
\begin{pmatrix}
i & 0 \\
0 & i
\end{pmatrix},\,\,\,\,
\bss{\si}^9 = 
\begin{pmatrix}
1 & 0 \\
0 & -1
\end{pmatrix},
}
$a=0,\ldots,7$ and $(\bss{\g}^{4},\bss{\g}^{5},\bss{\g}^{6},\bss{\g}^{7})=(\bs{e}_2,\bs{e}_7,\bs{1},\bs{e}_1)$. Then $\det(V) = \eta^{AB} p_A p_B$ with $\eta^{AB}$ the 10D Minkowski metric, so $\xi_1, \xi_2$ span the plane of the extra directions $x^8$ and $x^9$. For the most general $\Psi$, note from (\ref{Xi4D}) that
\eq{\label{realCond}
x_2 = x^{\ddagger}_1 \circ \bss{\g}_E^0
}
where $\ddagger$ combines operator adjoint, Hermitian adjoint and octonion conjugation (complex conjugation on numbers). Equivalently  $x_1 = \bss{\g}_E^0 \circ x^{\ddagger}_2$, so the condition reads $\tau(\Psi) = \Psi$, with the antilinear map $\tau(x_1,x_2) = (\bss{\g}_E^0 \circ x^{\ddagger}_2,x^{\ddagger}_1 \circ \bss{\g}_E^0)$. Since $\tau^2 = 1$, this is a reality condition and a general $\Psi$ reads $\Psi = \bss{\chi} + \tau(\bss{\chi})$ with $\bss{\chi} = (\bar{x}^{\mathbb{H}}_2, x^{\mathbb{H}}_1)^T$ and 
\ea{
x^{\mathbb{H}}_1 & = \chi^{(1)}_{1}\,\omega_2-\chi^{(1)}_{2}\,\omega_1+\chi^{(3)}_{1}\,\omega^\dagger_1+\chi^{(3)}_{2}\,\omega^\dagger_2\,, \label{x1H} \\
x^{\mathbb{H}}_2 & =\chi^{(2)}_{1}\,\omega_1-\chi^{(2)}_{2}\,\omega^\dagger_2+\chi^{(4)}_{1}\,\omega_2+\chi^{(4)}_{2}\,\omega^\dagger_1\,, \label{x2H}
}
with $x^{\mathbb{H}}_{1,2}\in \mathbb{C} \otimes \mathbb{H}$, so the four fields $\chi^{(g)} = (\chi^{(g)}_1,\chi^{(g)}_2)$ $(g=1,\ldots,4)$, defined in this way for future convenience, are all left-handed. Since $\bss{\chi}$ and $\tau(\bss{\chi})$ span $(\mathbb{C}\otimes \mathbb{H})^2$ and $(\mathbb{C}\otimes \mathbb{H}^{\perp})^2$, Eqs.~(\ref{x1H})-(\ref{x2H}) cover the 16 real parameters of $\Psi$. The 10D generalization of Eq.~(\ref{Xi4D}) is
\eq{\label{Weyl10DL}
\Xi^{(10)} =
\begin{pmatrix}
\bss{\si}^A \dd_{A} & \Psi \\
\bar{\Psi}^T & 0      
\end{pmatrix}
}
with Lagrangian
\eq{\label{10DL}
\ma{L}^{(10)} = \det(\Xi^{(10)}) = \Psi^{I} C^{A}_{IJ} \dd_{A}\Psi^{J}
}
where $C^{A}_{IJ} = -\textrm{Re}[\bar{\bs{e}}^{T}_I \tilde{\bss{\si}}^A \bs{e}_{J}]$ for a basis $\bs{e}_I$ of $\Psi = \Psi^I \bs{e}_I$ with $\tau(\bs{e}_I) = \bs{e}_I$, the derivative being ordered as in 4D. The $C^A$ generate $\ma{C}\ell_{1,9}$ via $\Gamma^A=\begin{psmallmatrix}0 & C^A \\ \eta^{AA} (C^A)^{-1} & 0\end{psmallmatrix}$ (no sum), so Eq.~(\ref{10DL}) is the 10D Majorana-Weyl Lagrangian in exceptional Jordan form, the highest dimensional one fitting in this algebra. That the cubic form of $J_3(\mathbb{C}\otimes\mathbb{O})$ contains the $\textrm{Spin}(1,9)$-invariant vector-spinor-spinor pairing is known~\cite{Sudbery:1984, Schray:1996, Baez:2010}. To our knowledge, its identification with the free Majorana-Weyl Lagrangian and the exact embedding of the 4D Dirac Lagrangian via Eq.~(\ref{g0gmuCl}) are new.

\section{Spacetime symmetries}

Equation (\ref{JordanL}) embeds, from first principles, the symmetries of the free Dirac Lagrangian into those of $J_3(\mathbb C\otimes\mathbb O)$, the determinant-preserving group $E_6(\mathbb C)$, under which $X$ is a $\mathbf{27}$. To act on fields, $g\in E_6(\mathbb{C})$ must not mix the spinor sector $(x_1,x_2)$ with $(V,\xi_3)$. These sectors form the Peirce decomposition $J_{\la}(\Pi) = \{X: \Pi \odot X = \la X\}$, $\la \in \{0, \frac{1}{2}, 1\}$ with $\Pi = \textrm{diag}(0,0,1)$, such that $J_{0}(\Pi)=V \oplus 0$, $J_{1/2}(\Pi)=\begin{psmallmatrix} 0 & \Psi \\ \bar{\Psi}^T & 0 \end{psmallmatrix}$ and $J_{1}(\Pi)=\textrm{diag}(0,0,\xi_3)$. The involution $\si_\Pi(X)=(1-2\Pi)X(1-2\Pi)$ is $+1$ on $J_0$, $J_1$ and $-1$ on $J_{1/2}$, so the $g$ commuting with it preserve each space, reducing $E_{6}(\mathbb{C})$ to $(\textrm{Spin}(10, \mathbb{C})\times \mathbb{C}^*)/\mathbb{Z}_4$ \cite{Yokota:1990} (with $\mathbb{C}^*= \mathbb{C}\backslash\{0\}$). Here, $\al \in \mathbb{C}^*$ acts as $X \mapsto D_{\la} X D_{\la}$, with $D_{\la} = \textrm{diag}(\la^{-1},\la^{-1},\la^{2})$ ($\la\in \mathbb{C}^*$), and $\mathbb{Z}_4$ consists of the pairs $(i^k,z^k)$, which act trivially, with $z$ the central element of $\textrm{Spin}(10, \mathbb{C})$ acting as $-i$ on the $\textbf{16}$. The $\mathbb{C}^*$ factor is removed by the reality condition. To see this, note that when extended to $X$ so that $\det(\tau(X))=(\det(X))^*$, $\tau$ reads $\tau(X) = D(P X^* P)\bar{D}^T$, with $D = \textrm{diag}(-\bss{\g}_E^0,\bss{\g}_E^0,\bs{1})$, $X^*$ the entrywise complex conjugate and $P$ exchanging rows and columns 1 and 2. Since $\tau\circ D_{\la} = D_{\la^*}\circ\tau$, preserving the Majorana condition requires $\la\in\mathbb{R}^*$. Real $\la$ acts as the coordinate rescaling $x\mapsto\la^2x$ with $\Psi\mapsto\la\Psi$, leaving the Lagrangian invariant but scaling the action as $S\mapsto\la^{20}S$, so it is no symmetry unless $\la^2=1$, and $\la=-1$ is the central element $z^2$. This leaves $\textrm{Spin}(10,\mathbb{C})$, the rotations of $\Xi^{(10)}$ leaving the action of Eq.~(\ref{10DL}) invariant, under which $\bs{27} \to \bs{1} \oplus \bs{10} \oplus \bs{16} (J_1 \oplus J_0 \oplus J_{1/2})$. Finally, $\tau(V(p)) = -V(p^*)$, so the reality condition requires purely imaginary $p_A$ and only the group $\textrm{Spin}(1,9)$ survives.

Independently of the underlying mechanism, reducing to 4D requires transformations that do not mix the four observed directions $\bss{\si}^{\mu}$ with the other six. Ignoring the reality condition, these are the $g \in \textrm{Spin}(10, \mathbb{C})$ commuting with the entry-wise involution $\theta(a) = -a$ for $a \in \mathbb{H}^{\perp}$ and $\theta(a) = a$ otherwise. Since $\theta=-\hat{\gamma}^5$ (cf. Appendix~B), this is the chirality operator of the Dirac theory itself. Commuting with $\theta$ selects ${(\textrm{Spin}(4, \mathbb{C}) \times \textrm{Spin}(6, \mathbb{C}))/\mathbb{Z}_2}$, with ${\bs{10}\to (\bs{2},\bs{2},\bs{1})\oplus (\bs{1},\bs{1},\bs{6})}$ and ${\bs{16}\to (\bs{2},\bs{1},\bs{4})\oplus (\bs{1},\bs{2},\bar{\bs{4}})}$, which the reality condition turns into $(\textrm{SL}(2, \mathbb{C}) \times SU(4))/\mathbb{Z}_2$, with ${\bs{16}\to (\bs{2},\bs{4})\oplus (\bar{\bs{2}},\bar{\bs{4}})}$. The 10D group thus reduces to the Lorentz group times an extra-dimensional $SU(4)$, seen in 4D as an internal symmetry, rotating $(\chi^{(1)},\chi^{(2)},\chi^{(3)},\chi^{(4)})^T$ as a $\bs{4}$ and the 10D Lagrangian decomposes as
\eq{\label{L10D4D}
\ma{L}^{(10)} = \sum_g \chi^{(g)\dagger}\,i\bar\sigma^\mu \dd_{\mu} \chi^{(g)}+\Big[\frac{1}{2}(\Sigma^a)_{gh} \chi^{(g)} \dd_a \chi^{(h)}+\text{h.c.}\Big]\,,
}
where the antisymmetric $\Sigma^a$ (read off from $C^a$) form the $\bs{6}$ of $SU(4)$ and $\g^0 \g^{\mu} = \textrm{diag}(\bar{\si}^\mu,\si^\mu)$. With full 10D dependence, the extra-dimensional momenta turn the second term into a mass term, pairing the four Weyl fermions into two towers of massive Dirac fermions. If the extra-dimensional geometry allows zero modes, four identical massless Weyl fermions remain, which a 4D observer would interpret as four families. $SU(4)$ is then naturally a family symmetry, a hypothesis constraining the extra dimensions once confronted with the observed families.

All evidence points to three chiral families, with a fourth chiral family being essentially excluded. We now show that this can be encoded in a complex structure $K$ on the extra dimensions. The Dirac embedding already fixes one in $\mathbb{C}\otimes\mathbb{O}$, namely $J=-\bs{e}_1$, the unique imaginary octonion with $J\circ\om^\dagger_\al = i\om^\dagger_\al$, which also satisfies $J\circ\om_\al=-i\om_\al$. We thus require $K$ to act only on $\bss{\si}^4,\dots,\bss{\si}^9$, to agree with $J$ on the four octonionic directions and to square to $-1$. Since $J\circ\bss{\g}^m=K^m{}_n\,\bss{\g}^n$ $(m,n=4,\dots,7)$, this fixes $K^4{}_5=-K^5{}_4=1$ and $K^7{}_6=-K^6{}_7=1$, while the antisymmetry, required to preserve the metric, fixes $K^8{}_9=-K^9{}_8=1$ (up to sign). Requiring $\de_K\ma{L}^{(10)}=0$ then uniquely gives $\de_K\chi^{(g)}=\la^{g}{}_{f}\chi^{(f)}$, with $\la=\frac{i}{2}\textrm{diag}(1,1,1,-3)$ (cf. Appendix~C), together with $\de_K\dd_a=K^b{}_a\dd_b$, so $K$ rotates simultaneously the planes $(x^4,x^5)$, $(x^6,x^7)$ and $(x^8,x^9)$.

Now take the extra dimensions to be a flat torus $T^6 = \mathbb{R}^6/\Lambda: y \sim y+\ell\,\, (\ell \in \Lambda)$, where $\Lambda$ is a hexagonal lattice in each $K$-plane, and with the orbifold twist $T^6 \to T^6/\mathbb{Z}_6\,, \quad y \sim e^{-\frac{\pi}{3}K} y$ where $y^n \equiv x^n$ $(n=4,\ldots,9)$. Testing the family interpretation requires an internal space $\ma{W}$ with basis $\bss{\e}_{\dot{r}}$, $\dot{r}=1,\ldots,N$, with $\sum_{\dot{r}}\Psi_{\dot{r}} \otimes \bss{\e}_{\dot{r}} \in J_{1/2}(\Pi) \otimes \ma{W}$ and Lagrangian $\ma{L}_{\textrm{ext}}=h_{\dot{r}\dot{s}}\Psi^{I\dot{r}} C^A_{IJ} \dd_A \Psi^{J\dot{s}}$. Only a real, symmetric and positive $h_{\dot{r}\dot{s}}$ contributes without ghosts, so $h_{\dot{r}\dot{s}}=\de_{\dot{r}\dot{s}}$ and the transformations of $\dot{r}$ are $O(N)$. The orbifold then identifies $\Psi_{\dot{r}}(x,Py)=\rho_{\dot{r}}{}^{\dot{s}}\,\hat{P}\,\Psi_{\dot{s}}(x,y)$ where $P\equiv e^{-\frac{\pi}{3}K}$, $\hat{P}$ is its action on the fields ($\chi^{(g)}\to e^{\frac{\pi}{3}\la_g}\chi^{(g)}$), and $\rho \in O(N)$. In fact, writing $\Psi_{\dot{r}} \equiv \Psi(\chi_{\dot{r}})$, we can use that $\rho$ is real to extend the orbifold map to the fields $\chi^{(g)}_{\dot{r}}$ as $\chi^{(g)}_{\dot{r}}(x,Py)=e^{\frac{\pi}{3}\la_g}\rho_{\dot{r}}{}^{\dot{s}}\,\chi^{(g)}_{\dot{s}}(x,y)$. Since $P^6=1$ but $\hat{P}^6=e^{2\pi\la}=-1$, consistency requires $\rho^6=-1$, so $\ma{J}\equiv-\rho^3$ is a complex structure on $\ma{W}$ and $\dot{r}$ runs over $2n$ values, carrying a representation $R\oplus\bar{R}$, with $R\simeq\mathbb{C}^n$ the $\ma{J}=+i$ eigenspace of $\ma{W}\otimes\mathbb{C}$, $i$ being the imaginary unit of $\chi^{(g)}$ and $\la$.

 In a basis $\dot{I}=1,\ldots,n$ of $R$ diagonalizing $\rho$, $\rho^3=-\ma{J}=-i$ gives eigenvalues $z_{\dot{I}} \in \{e^{-i\pi/6},e^{i\pi/2},e^{-i5\pi/6}\}$ on $R$ and $z_{\dot{I}}^*$ on $\bar{R}$. If the orbifold preserves $g_{\rm int}\in U(n)$, $\rho$ commutes with it and, since $R$ is irreducible, $z_{\dot{I}}\equiv z$ for all $\dot{I}$. With the eigencombinations $\chi^{(g)}_{\dot{I},\pm}$, such that $\rho \chi^{(g)}_{\dot{I},+} = z \chi^{(g)}_{\dot{I},+}$ and $\rho \chi^{(g)}_{\dot{I},-} = z^{*} \chi^{(g)}_{\dot{I},-}$, the orbifold reads
\eq{\label{OrbiPhase}
\chi^{(g)}_{\dot{I},+}(x,Py)=e^{\frac{\pi}{3}\la_g}z\,\chi^{(g)}_{\dot{I},+}(x,y)
}
and similarly for $\chi^{(g)}_{\dot{I},-}$ with $z \to z^*$. Massless modes, i.e. light families, are $y$-independent and thus need orbifold phase 1.
\begin{table}[h!]
\centering
\setlength{\tabcolsep}{10pt}
\renewcommand{\arraystretch}{1.3}
\begin{tabular}{c|c|cc}
\hline\hline
\multirow{2}{*}{rep} & \multirow{2}{*}{$\rho$ eigenvalue} & \multicolumn{2}{c}{Orbifold phase} \\
\cline{3-4}
    &         & $g=1,2,3$ & $g=4$ \\
\hline
          & $e^{-i\pi/6}$  & $1$             & $e^{-2\pi i/3}$ \\
$R$       & $e^{i\pi/2}$   & $e^{2\pi i/3}$  & $1$ \\
          & $e^{-5i\pi/6}$ & $e^{-2\pi i/3}$ & $e^{2\pi i/3}$ \\
\hline
          & $e^{i\pi/6}$   & $e^{i\pi/3}$    & $e^{-i\pi/3}$ \\
$\bar{R}$ & $e^{-i\pi/2}$  & $e^{-i\pi/3}$   & $-1$ \\
          & $e^{5i\pi/6}$  & $-1$            & $e^{i\pi/3}$ \\
\hline\hline
\end{tabular}
\caption{Orbifold phases $e^{\frac{\pi}{3}\lambda_g}z$ of $\chi^{(g)}_{\dot{I},+}$ ($R$) and $e^{\frac{\pi}{3}\lambda_g}z^*$ of $\chi^{(g)}_{\dot{I},-}$ ($\bar R$), for each possible eigenvalue of $\rho$.}
\label{tab:OrbiPhase}
\end{table}
By Table~1, only two choices of $z$ give massless modes: only $\chi^{(4)}_{\dot{I},+}$ ($z=e^{i\pi/2}$), giving one light family, or $\chi^{(1,2,3)}_{\dot{I},+}$ ($z=e^{-i\pi/6}$), giving three, only the latter being consistent with experiment. Reversing the sign of the $(8,9)$ block of $K$ maps $\la$ to $\frac{i}{2}\textrm{diag}(-1,3,-1,-1)$ and exchanges $R$ and $\bar{R}$, leaving this count unchanged. The extra-dimensional momentum term $p_{a}\Sigma^{a}_{gh}$ acts as a mass term which, since $h^{\dot{I}\dot{J}}=\de^{\dot{I}\dot{J}}$ pairs $R$ with $\bar{R}$ and $\Sigma^{a}$ is antisymmetric, couples $\chi^{(4)}_{\dot{I},+}$ to $\chi^{(1,2,3)}_{\dot{I},-}$. The resulting Dirac fermion has left-handed part $\chi^{(4)}_{\dot{I},+}$ and right-handed part $\chi^{(h)\dagger}_{\dot{I},-}$, both in $R$, so the fourth family is a vector-like KK excitation at the compactification scale, evading the exclusion of a chiral one. This scale also contains vector-like copies of $\chi^{(1,2,3)}_{\dot{I},+}$, with mirror fields $\chi^{(h)\dagger}_{\dot{I},-}$ as right-handed partners. Crucially, no massless mirror family arises for any allowed $\rho$, as no entry of the $\bar{R}$ block equals $1$. $SU(4)_F$ is therefore a family symmetry, broken by the orbifold to $SU(3)_F\times U(1)_F$ on the zero modes.

Independently of the internal space, $R$ and $\bar{R}$ are related by complex conjugation, with real structure $\kappa(\bss{\e}_{\dot{I}}) = \bss{\e}^*_{\dot{I}}$. For $\Theta = \sum \Psi^{I \dot{I}} \, \bs{e}_{I} \otimes \bss{\e}_{\dot{I}}$, the conjugate is $\Theta^c = (\tau \otimes \kappa)\Theta$. With $\bar{\Theta}$ the octonion conjugate and $1_{\textrm{int}}$ the internal identity, we define
\eq{
\Xi_{\textrm{ext}}=
\begin{pmatrix}
\bss{\si}^A \dd_A \otimes 1_{\textrm{int}} & \Theta \\ 
\bar{\Theta}^{c\,T} & 0
\end{pmatrix}\,,
}
so that $\ma{L}_{\textrm{ext}} = \det(\Xi_{\textrm{ext}})=-\textrm{Re}[\bar{\Theta}^{c\,T}(\tilde{\bss{\si}}^A \dd_A \otimes 1_{\textrm{int}})\Theta]=h(\bss{\e}_{\dot{I}},\bss{\e}_{\dot{J}}) \, \Psi^{I\dot{I}} C^A_{IJ} \dd_A \Psi^{J\dot{J}}$, with $h(\bss{\e}_{\dot{I}},\bss{\e}_{\dot{J}}) = \bss{\e}^{* T}_{\dot{I}} \bss{\e}_{\dot{J}}=\de_{\dot{I}\dot{J}}$. Preserving this bilinear and the orbifold condition restricts transformations of $\Xi_{\textrm{ext}}$ to $g = (g_{\textrm{ext}},g_{\textrm{int}})$ where ${g_{\textrm{ext}} \in \textrm{Spin}(1,3)\times SU(3)_F \times U(1)_F}$ acts on external indices and $g_{\textrm{int}} \in U(n)$ on internal ones. The family symmetry must eventually be broken to allow family mixing.

\section{Internal symmetries}

The internal sector follows simply by mimicking the external one. We take $\bss{\e}_{\dot{I}} = \bs{e}_{\dot{I}}$, with $\dot{I}=1,\ldots,16$ matching as the external spinor index and define
\eq{
\Xi_{\textrm{int}}=
\begin{pmatrix}
H_{\dot{A}} \otimes \bss{\si}^{\dot{A}} & \Theta \\
\bar{\Theta}^{T} & 0
\end{pmatrix}
}
where $H_{\dot{A}}$ may act on the external factor. As before, we obtain the internal Lagrangian $\ma{L}_{\textrm{int}} = \det(\Xi_{\textrm{int}})+\textrm{c.c.} = \Psi^{I\dot{I}} (H_{\dot{A}})_{IJ} G^{\dot{A}}_{\dot{I}\dot{J}}\Psi^{J\dot{J}}+\textrm{c.c.}$, where ${(H_{\dot{A}})_{IJ} = \textrm{Re}[\bar{\bs{e}}^T_I H_{\dot{A}} \bs{e}_J]}$, ${G^{\dot A}_{\dot I\dot J} = -\textrm{Re}[\bar{\bss{\e}}^T_{\dot{I}}\tilde{\bss{\si}}^{\dot{A}}\bss{\e}_{\dot{J}}]}$ and the $H_{\dot{A}}$ must then contain bosonic fields only. Since $G^{\dot{A}}_{\dot{I}\dot{J}}$ is symmetric, $(H_{\dot{A}})_{IJ}$ is an antisymmetric $16 \times 16$ matrix, i.e. the operator $H_{\dot{A}}=-\frac{1}{{3!}}\,H_{ABC\dot{A}}\tilde{\bss{\si}}^{[A}\Big(\bss{\si}^{B}\Big(\tilde{\bss{\si}}^{C]}\cdot\Big)\Big)$, with components $(H_{\dot{A}})_{IJ} = \frac{1}{3!}H_{ABC \dot{A}} C^{ABC}_{IJ}$, where $C^{ABC}_{IJ} = \big(C^{[A}\,(\eta^{BB}C^{B})^{-1}\,C^{C]}\big)_{IJ}$. We now hypothesize that $\Xi_{\textrm{int}}$ preserves the same involutions as its external counterpart, although in compact form since $g_{\textrm{int}} \in U(n)$. This gives the chain
\ea{
E_6 &\xrightarrow{\makebox[1.1em]{$\scriptstyle\sigma_\Pi$}} \textrm{Spin}(10)\!\times\! U(1)_{\Pi} \nonumber\\
    &\xrightarrow{\makebox[1.1em]{$\scriptstyle\theta$}} SU(2)_L\!\times\! SU(2)_R\!\times\! SU(4)_C\!\times\! U(1)_{\Pi} \nonumber\\
    &\xrightarrow{\makebox[1.1em]{$\scriptstyle K$}} SU(2)_L\!\times\! SU(2)_R\!\times\! SU(3)_C\!\times\! U(1)_{B-L}\!\times\! U(1)_{\Pi}\,, \nonumber
}
omitting global quotients. Unlike externally, where the reality condition and the invariance of the action remove the Peirce factor, the internal fields obey no such condition and $V_{\textrm{int}}$ is a dynamical field rather than a derivative, so the phases $|\la|=1$ preserve $\ma{L}_{\textrm{int}}$ and survive as a $U(1)_\Pi$. The orbifold twist $\rho=e^{-i\pi/6}$ is an element of $U(1)_\Pi$, whose charges conveniently describe both fermion and scalar twist phases. Whether $U(1)_\Pi$ should be gauged is open. The group $[\mathrm{Spin}(10)\times U(1)_\Pi]/\mathbb Z_4$ is the stabilizer in Boyle’s construction \cite{Boyle:2020}, here externally as a consequence of the Dirac Lagrangian, and the internal role of $K$ parallels his splitting $\mathbb{O} = \mathbb{C} \oplus \mathbb{C}^3$, here of external origin. The color embeddings, however, differ in these Jordan coordinates. Boyle’s $SU(3)_C$ acts on $\mathbb C^3$ and fixes $\operatorname{diag}(1,-1,0)$, which here is the $\bss{\si}^{9}$ direction in the $\bs{6}$ of $SU(4)_C$, branching as $\bs{3} \oplus \bar{\bs{3}}$ with no color singlet. Thus the two reductions reproduce the same quark–lepton branching of the $\bs{16}$ through distinct actions on the Jordan vector sector. 

The internal components of $\Theta$ transform as $\bs{16}_1\to(\bs{2},\bs{1},\bs{3})_{\frac13}\oplus(\bs{2},\bs{1},\bs{1})_{-1}\oplus(\bs{1},\bs{2},\bar{\bs{3}})_{-\frac13}\oplus(\bs{1},\bs{2},\bs{1})_{1}$, a full SM family plus a right-handed neutrino, while $H_{ABC\dot A}$ is a 10D three-form in the $\bs{10}_{-2}$, with $\bs{10}\to(\bs{2},\bs{2},\bs{1})_0\oplus(\bs{1},\bs{1},\bs{3})_{-\frac23}\oplus(\bs{1},\bs{1},\bar{\bs{3}})_{\frac{2}{3}}$. The internal twist gives $H$ the phase $e^{i\pi/3}$, so $H$ is complex. With the family charge normalized by $\bs{4}=\bs{3}_1\oplus\bs{1}_{-3}$, a three-form component of family charge $q$ has total phase $e^{i\pi/3}e^{i\pi q/6}$. Among the charges $q=-6,-2,2,6$ of the purely extra-dimensional three-form, only $q=-2$ has a constant mode, in $\Lambda^2\bs{3}\otimes\bar{\bs{3}}=\bar{\bs{6}}\oplus\bs{3}$. Since the light-family bilinear is a $\bs{6}_2$, only the $\bar{\bs{6}}_{-2}$ couples to it. The same rule gives constant modes of the mixed components $H_{\mu\nu a\dot A}$ with $a$ in the $\bs{3}_{-2}$ of $\bs{6}=\bar{\bs{3}}_{2}\oplus\bs{3}_{-2}$, giving 4D antisymmetric tensors in the $(\bs{3},\bs{10}_{-2})$ that couple to the light-family tensor bilinears $\chi^{(g)}\sigma^{\mu\nu}\chi^{(h)}$, antisymmetric in $g,h$, whose propagation depends on the yet-underived kinetic term of $H$. No independent family-indexed Yukawa matrices are introduced. Flavour dependence enters through the bidoublet vacuum expectation values $\kappa_{u,d}^{gh}$, giving symmetric mass matrices $M_{u,d}^{gh}\propto\kappa_{u,d}^{gh}$.

Several aspects remain under development. A scalar in the $\mathbf{10}$ can  neither break the left-right symmetric group to the SM, nor give the right-handed neutrinos Majorana masses, calling for an additional $B-L$-breaking sector. The kinetic terms of the scalar and tensor sectors, as well as the scalar potential whose vacuum fixes the family structure of the fermion masses, also need to be derived. A mechanism is also required to separate the Higgs doublets from their colour-triplet partners. Other Jordan invariants in the Lagrangian may address these points. The anomalies of $U(1)_{\Pi}$, if gauged, and the phenomenology of the vector-like KK spectrum require a dedicated analysis as well. Finally, if $\textrm{Spin}(10)$ is gauged in the bulk, a second Majorana–Weyl spinor $\Psi_-$ in the same internal representation as $\Psi$ and of opposite 10D chirality would cancel the perturbative anomaly polynomial in the smooth bulk. To avoid introducing additional fermion zero modes, one may impose the orbifold boundary condition $\Psi_-(x,Py)=\hat{P}_-\rho D_-\Psi_-(x,y)$, where $\hat{P}_-$ denotes the spin lift on the opposite-chirality field. The additional order-three factor $D_-=\exp[-(2\pi/3)\ma{J}]$ projects out all constant modes of $\Psi_-$ in the free theory. Whether this relative phase follows from the present algebraic structure or requires an extension remains open. Its compatibility with the interactions and the cancellation of any anomalies localized at the orbifold fixed loci also remain to be established.

\begin{acknowledgments}
The author thanks the hospitality of the Maryland Center for Fundamental Physics where part of this research was developed as a visiting researcher, supported by the grant FCT-Mobility/1312232346/2024-25.
\end{acknowledgments}


\clearpage

\section*{Appendix}
\appendix

\section*{Appendix A: Derivation of Eq.~(\ref{g0gmuOct})}

\begin{figure}[h!]
  \centering
  \includegraphics[width=0.5\columnwidth]{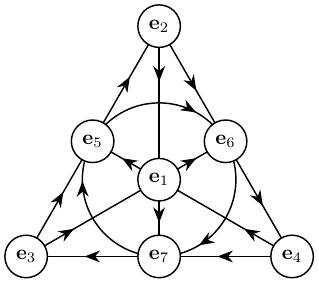}
  \caption{Fano plane summarizing the multiplication table of $\mathbb{O}$ adopted in this work. The six straight lines and the circle define the product ordering. Reversing the order introduces a minus sign. For example, $\bs{e}_3 \circ \bs{e}_5 = \bs{e}_2$ and $\bs{e}_2 \circ \bs{e}_5 = -\bs{e}_3$.}
  \label{fano}
\end{figure}

Consider the following generators $e_k$ of $\ma{C}\ell_{8,0}$ given by
\eq{\label{eAbasis}
\sum_{k=1}^{8} c_k e^k = \sigma^+\otimes A + \sigma^-\otimes B ,
}
where $\sigma^+=\left(\begin{smallmatrix}0&1\\0&0\end{smallmatrix}\right)$,
$\sigma^-=\left(\begin{smallmatrix}0&0\\1&0\end{smallmatrix}\right)$ and
{\setlength{\arraycolsep}{2pt}
\ea{
A&=\begin{pmatrix}
 c^{++}_{72} & 0 & c^{+-}_{54} & c^{-+}_{63} & c^{-+}_{81} & 0 & 0 & 0 \\
 c^{--}_{81} & 0 & 0 & 0 & c^{-+}_{72} & 0 & c^{+-}_{54} & c^{-+}_{63} \\
 c^{--}_{54} & c^{-+}_{63} & c^{+-}_{72} & 0 & 0 & 0 & c^{-+}_{81} & 0 \\
 c^{++}_{63} & c^{-+}_{54} & 0 & c^{+-}_{72} & 0 & 0 & 0 & c^{-+}_{81} \\
 0 & c^{++}_{72} & c^{++}_{63} & c^{++}_{54} & 0 & c^{+-}_{81} & 0 & 0 \\
 0 & c^{--}_{81} & 0 & 0 & 0 & c^{+-}_{72} & c^{++}_{63} & c^{++}_{54} \\
 0 & 0 & c^{--}_{81} & 0 & c^{--}_{54} & c^{+-}_{63} & c^{--}_{72} & 0 \\
 0 & 0 & 0 & c^{--}_{81} & c^{++}_{63} & c^{+-}_{54} & 0 & c^{--}_{72}
\end{pmatrix},\\
B&=\begin{pmatrix}
 c^{+-}_{72} & c^{-+}_{81} & c^{-+}_{54} & c^{+-}_{63} & 0 & 0 & 0 & 0 \\
 0 & 0 & c^{--}_{63} & c^{--}_{54} & c^{+-}_{72} & c^{-+}_{81} & 0 & 0 \\
 c^{++}_{54} & 0 & c^{++}_{72} & 0 & c^{+-}_{63} & 0 & c^{-+}_{81} & 0 \\
 c^{--}_{63} & 0 & 0 & c^{++}_{72} & c^{+-}_{54} & 0 & 0 & c^{-+}_{81} \\
 c^{--}_{81} & c^{--}_{72} & 0 & 0 & 0 & 0 & c^{-+}_{54} & c^{+-}_{63} \\
 0 & 0 & 0 & 0 & c^{++}_{81} & c^{++}_{72} & c^{++}_{63} & c^{++}_{54} \\
 0 & c^{++}_{54} & c^{--}_{81} & 0 & 0 & c^{+-}_{63} & c^{-+}_{72} & 0 \\
 0 & c^{--}_{63} & 0 & c^{--}_{81} & 0 & c^{+-}_{54} & 0 & c^{-+}_{72}
\end{pmatrix}.
}
}
where we defined $c^{s_1  s_2}_{kj} \equiv s_1 c_k+s_2 i c_j$. Consider now the identity in Eq.~(\ref{g0gmuCl}) where $u^{\mu} =\exp\left[i \frac{\pi}{2} (\g^0\g^{\mu})_{\al\be} \om^{\al \dagger}\om^{\be}\right]$, $w = e_{1234} + e_{1467} - e_{1357} + e_{1256} - e_{3456}  - e_{2457} - e_{2367}$ and $I_e = e_1 e_2 \ldots e_8$ (as derived in \cite{Quinta:2025}). For $x \in \mathbb{C} \otimes \ma{C}\ell_{8,0}$, let $\braket{x}_1$ denote the component of $x$ made of linear combinations of $e_k$. Then $\braket{x}_1 = \textrm{Tr}[x e_k] e_k/16$ and Eq.~(\ref{g0gmuCl}) can be alternatively written as
\eq{\label{g0gmuinner}
i(\g^0\g^{\mu})_{\al\be} = 2 \om_{\al} \cdot \braket{u^{\mu}\om^{\dagger}_{\be} (1+w)(1+I_e)}_1
}
where $(x \cdot y) = x^i y_i$ is the inner product (without conjugation). Defining
\eq{
n^{i,j}_{\pm} = \frac{e_i \pm e_j}{\sqrt{2}}\,,
}
one can express the spinors $u^{\mu} =\exp\left[i \frac{\pi}{2} (\g^0\g^{\mu})_{\al\be} \om^{\al \dagger}\om^{\be}\right]$, in the Weyl basis, as the following products
\ea{
u^{0} & = -e_1 n^{1,8}_- e_2 n^{2,7}_- e_3 n^{3,6}_- e_4 n^{4,5}_- \,, \label{u0} \\
u^{1} & = e_1 n^{1,7}_+ e_2 n^{2,8}_+ e_3 n^{3,5}_- e_4 n^{4,6}_- \,, \label{u1} \\
u^{2} & = e_1 n^{1,2}_- e_7 n^{7,8}_+ e_3 n^{3,4}_+ e_5 n^{5,6}_- \,, \label{u2} \\
u^{3} & = e_1 n^{1,8}_+ e_2 n^{2,7}_- e_3 n^{3,6}_- e_4 n^{4,5}_+ \,. \label{u3} 
}
Now recall that $a,b \in \mathbb{R}^8$ are a representation of octonions with a product defined in terms of the calibration $w$ and neutral axis $n$ via
\ea{\label{OctoProd}
a \circ b & = \braket{anb(1+w)(1-I_e)}_1\,,
}
leading to the octonion identification $\textbf{e}_k=e_k$ $(k=1,\ldots,7)$ and $\textbf{1} = e_8$ used throughout this work, with the corresponding Fano plane given by Fig.~\ref{fano}.
The real part of an octonion in $\ma{C}\ell_{8,0}$ is identified with $\textrm{Re}[x] \equiv x \cdot n$ and the octonion conjugate by $\bar{x} = 2\textrm{Re}[x] - x = 2(x \cdot n)n-x$, implying in particular that
\eq{\label{xdoty}
x \cdot y = \textrm{Re}[\bar{x} \circ y]\,.
}
Using the octonion identification in $\ma{C}\ell_{8,0}$, combining Eqs.~(\ref{u0})-(\ref{u3}) with Eq.~(\ref{OctoProd}) and using the fact that $a \circ b = \braket{bna(1+w)(1+I_e)}_1$, we find that
\eq{\label{uMmu}
\braket{u^{\mu} \om^{\dagger}_{\be} (1+w)(1+I_e)}_1 = M^{\mu}(\om^{\dagger}_{\be})
}
where $M^{\mu}(x)$ are the octonionic functions defined via
\begin{widetext}
\ea{
M^{0}(x) & = -((((((((x \circ \bar{n}^{4,5}_-) \circ e_4) \circ \bar{n}^{3,6}_-) \circ e_3) \circ \bar{n}^{2,7}_-) \circ e_2) \circ \bar{n}^{1,8}_-) \circ e_1) = -e_1 \circ x \,, \\
M^{1}(x) & = ((((((((x \circ \bar{n}^{4,6}_-) \circ e_4) \circ \bar{n}^{3,5}_-) \circ e_3) \circ \bar{n}^{2,8}_+) \circ e_2) \circ \bar{n}^{1,7}_+) \circ e_1) = -(e_5 \circ ( e_3 \circ x))\,, \\
M^{2}(x) & = ((((((((x \circ \bar{n}^{5,6}_-) \circ e_5) \circ \bar{n}^{3,4}_+) \circ e_3) \circ \bar{n}^{7,8}_+) \circ e_7) \circ \bar{n}^{1,2}_-) \circ e_1) = (e_5 \circ ( e_6 \circ x))\,, \\
M^{3}(x) & = ((((((((x \circ \bar{n}^{4,5}_+) \circ e_4) \circ \bar{n}^{3,6}_-) \circ e_3) \circ \bar{n}^{2,7}_-) \circ e_2) \circ \bar{n}^{1,8}_+) \circ e_1) = (e_5 \circ (e_4 \circ x))\,.
}
\end{widetext}
Since $M^{0}(\om^{\dagger}_{\be}) = (e_5 \circ ( (-ie_5) \circ \om^{\dagger}_{\be}))$, one can note a pattern in $M^{\mu}(x)$, namely that
\eq{
M^{\mu}(\om^{\dagger}_{\al}) = i \bss{\g}^0 \circ (\bss{\g}^{\mu} \circ  \om^{\dagger}_{\al})
}
after defining $(\bss{\g}^0,\bss{\g}^1,\bss{\g}^2,\bss{\g}^3) = (-i\bs{e}_5,-\bs{e}_3,\bs{e}_6,\bs{e}_4)$. Using the above equation in Eqs.~(\ref{uMmu}), inserting it in (\ref{xdoty}) and subsequently in (\ref{g0gmuinner}), we arrive at Eq.~(\ref{g0gmuOct}).

\section*{Appendix B: Left-handed and right-handed subspaces}

In the Weyl basis used in this work, the chirality matrix is
\eq{\label{g5Weyl}
\g^5 = i\g^0\g^1\g^2\g^3 = \textrm{diag}(-1_{2},\,1_{2})\,.
}
The left-handed components of $\psi$ are therefore $\psi_1,\psi_2$, selected by $P_L=(1-\g^5)/2$, and the right-handed ones are $\psi_3,\psi_4$. To translate this into octonionic language, note that Eq.~(\ref{g0gmuOct}), together with $2\,\om_{\al}\cdot\om^{\dagger}_{\be}=\de_{\al\be}$, states that the maps $M^{\mu}(x)=i\bss{\g}^0\circ(\bss{\g}^{\mu}\circ x)$ represent the matrices $i\g^0\g^{\mu}$ on the span of the $\om^{\dagger}_{\be}$, i.e.
\eq{\label{Mmatrix}
M^{\mu}(\om^{\dagger}_{\be}) = \sum_{\al} i(\g^0\g^{\mu})_{\al\be}\,\om^{\dagger}_{\al}\,.
}
Since $\g^5=(i\g^0\g^1)(i\g^0\g^2)(i\g^0\g^3)$, the chirality matrix is represented by
\ea{
\hspace{-0.4mm} \hat{\g}^5(x) &\equiv M^1\big(M^2\big(M^3(x)\big)\big) \nonumber\\
&= -e_5\circ\Big(e_3\circ\Big(e_5\circ\Big(e_6\circ\big(e_5\circ(e_4\circ x)\big)\Big)\Big)\Big)\,.
}
This expression simplifies with the left Moufang identity $a\circ(b\circ(a\circ x))=(a\circ b\circ a)\circ x$, valid in any alternative algebra. Since distinct imaginary units anticommute and square to $-1$, we have $e_5\circ e_6\circ e_5=-e_5\circ e_5\circ e_6=e_6$, and so
\eq{\label{G5oct}
\hat{\g}^5(x)=-e_5\circ\big(e_3\circ\big(e_6\circ(e_4\circ x)\big)\big)\,.
}
The operator $\hat{\g}^5$ now involves only left multiplications by the four units of $\mathbb{H}^{\perp}$. The decomposition $\mathbb{O}=\mathbb{H}\oplus\mathbb{H}^{\perp}$ is a $\mathbb{Z}_2$-grading, with $\mathbb{H}\circ\mathbb{H}\subset\mathbb{H}$, $\mathbb{H}\circ\mathbb{H}^{\perp}\subset\mathbb{H}^{\perp}$ and $\mathbb{H}^{\perp}\circ\mathbb{H}^{\perp}\subset\mathbb{H}$. Each of these multiplications therefore exchanges $\mathbb{H}$ and $\mathbb{H}^{\perp}$, so that $\hat{\g}^5$ maps $\mathbb{H}$ to $\mathbb{H}$ and $\mathbb{H}^{\perp}$ to $\mathbb{H}^{\perp}$. Its eigenvalues follow from the multiplication table of Fig.~\ref{fano}. For instance, for $x=\bs{1}$ we have $e_4\circ\bs{1}=e_4$, $e_6\circ e_4=e_2$, $e_3\circ e_2=-e_5$ and $e_5\circ(-e_5)=\bs{1}$, so $\hat{\g}^5(\bs{1})=-\bs{1}$, while for $x=e_3$ we have $e_4\circ e_3=-e_7$, $e_6\circ(-e_7)=-e_5$, $e_3\circ(-e_5)=-e_2$ and $e_5\circ(-e_2)=-e_3$, so $\hat{\g}^5(e_3)=e_3$. Repeating this for the remaining units gives
\eq{\label{G5theta}
\hat{\g}^5(x) =
\begin{cases}
-x\,, & x\in\mathbb{H}\,,\\
+x\,, & x\in\mathbb{H}^{\perp}\,.
\end{cases}
}
Since $\hat{\g}^5$ is linear over $\mathbb{C}$, the same holds on $\mathbb{C}\otimes\mathbb{H}$ and $\mathbb{C}\otimes\mathbb{H}^{\perp}$, i.e.\ $\hat{\g}^5=-\theta$, with $\theta$ the involution defined in the main text. The octonionic chirality projectors are then
\eq{\label{PLPR}
P_L=\frac{1-\hat{\g}^5}{2}=\frac{1+\theta}{2}\,,\qquad P_R=\frac{1+\hat{\g}^5}{2}=\frac{1-\theta}{2}\,,
}
which project onto $\mathbb{C}\otimes\mathbb{H}$ and $\mathbb{C}\otimes\mathbb{H}^{\perp}$, respectively. This agrees with the Witt basis, since $\om^{\dagger}_1=-(ie_1+\bs{1})/2$ and $\om^{\dagger}_2=-(ie_2+e_7)/2$ lie in $\mathbb{C}\otimes\mathbb{H}$, while $\om^{\dagger}_3=-(ie_3+e_6)/2$ and $\om^{\dagger}_4=-(ie_4+e_5)/2$ lie in $\mathbb{C}\otimes\mathbb{H}^{\perp}$. Hence $\psi_L=\psi_1\om^{\dagger}_1+\psi_2\om^{\dagger}_2\in\mathbb{C}\otimes\mathbb{H}$ and $\psi_R=\psi_3\om^{\dagger}_3+\psi_4\om^{\dagger}_4\in\mathbb{C}\otimes\mathbb{H}^{\perp}$.

Finally, $\hat{\g}^5$ is defined on the whole of $\mathbb{C}\otimes\mathbb{O}$, not only on the span of the $\om^{\dagger}_{\be}$, and there it commutes with the octonionic boost and rotation generators $x\mapsto -iM^k(x)$ and $x\mapsto M^l(M^m(x))$, which represent $\g^0\g^k$ and $\g^l\g^m$, respectively. Its eigenspaces are therefore Lorentz invariant. On all of $\mathbb{C}\otimes\mathbb{H}$ the boosts act as $-i$ times the rotations, i.e.\ $\g^0\g^k\to-i\,\g^l\g^m$ with $(k,l,m)$ cyclic, exactly as on $\psi_1,\psi_2$, while on $\mathbb{C}\otimes\mathbb{H}^{\perp}$ they act as $+i$ times the rotations. Every element of $\mathbb{C}\otimes\mathbb{H}$, including those built from $\om_1$ and $\om_2$, is thus left-handed, and every element of $\mathbb{C}\otimes\mathbb{H}^{\perp}$ is right-handed.

\section*{Appendix C: Action of $K$ on the fields}

To define $K$ on all of $\Xi^{(10)}$, its action on $\Psi$ must satisfy $\de_{K}\ma{L}^{(10)} = 0$. This is most easily achieved by taking the $\Psi\equiv\Psi(\chi^{(g)})$ decomposition, such that the action of $K$ becomes encoded through the fields $\chi^{(g)}$ as the matrix $\la$ such that $K(\Psi(\chi^{(g)})) = \Psi(\la^{g}{}_{f} \chi^{(f)})$. Using (\ref{L10D4D}), invariance of the kinetic term requires ${\de_K(\chi^{\dagger}\bar{\si}^{\mu} \dd_{\mu}\chi)=\chi^{\dagger}(\la^{\dagger}+\la)\bar{\si}^{\mu} \dd_{\mu}\chi=0}$, so $\la^{\dagger}=-\la$, while invariance of the extra-dimensional term requires $\de_K \Big(\frac{1}{2}\Sigma^a_{gh}\chi^{(g)}\dd_a\chi^{(h)}\Big)=\frac{1}{2}\big(\la^T\Sigma^a+\Sigma^a \la+K^a{}_b\,\Sigma^b\big)_{gh}\,\chi^{(g)}\dd_a\chi^{(h)}=0$, i.e. $\la^T\Sigma^a+\Sigma^a \la=-K^a{}_b\,\Sigma^b$ $(a=4,\dots,9)$. The latter equation, together with the anti-Hermitian condition, uniquely selects $\la = \frac{i}{2}\textrm{diag}(1,1,1,-3)$. The derivatives transform according to $\de_K \dd_{\mu}=0$ and $\de_K \dd_{a} = K^{b}{}_{a} \dd_b$, which induces the infinitesimal coordinate transformation $\de x^A=-K^{A}{}_B x^B$, with $x^\mu$ unchanged since $K^\mu{}_B=0$.

\end{document}